\documentclass[twoside]{article}
\usepackage{fleqn,espcrc2}
\usepackage{psfig}
\title{Novel features of $J/\Psi$ dissociation in matter}
\author{
 A. Sibirtsev$^{1,2}$, K. Tsushima$^1$, K. Saito$^{1,3}$ and  A. W. 
Thomas$^1$ \\ \vspace{3mm}
{$^1$Special Research Center for the Subatomic Structure of Matter 
and Department of Physics and Mathematical Physics,
University of Adelaide, SA 5005, Australia \\
$^2$ Institut f\"ur Theoretische Physik, Universit\"at Giessen,
D-35392 Giessen, Germany  \\  
$^3$ Physics Division, Tohoku College of Pharmacy, 
Sendai 981-8558, Japan}}
\begin{document}

\begin{abstract}
We make a detailed study of the effect that the recently 
predicted modification of the in-medium masses of charmed mesons 
would have on $J/\Psi$ dissociation on pion and $\rho$-meson
comovers in relativistic heavy ion collisions.
We find a substantial dependence of the $J/\Psi$ absorption 
rates on the density and temperature of the nuclear matter. 
This suggests that a quantitative analysis of
$J/\Psi$ dissociation in nucleus nucleus collisions
should include the effects of the modification of meson masses in
dense matter. 
%
\end{abstract}
\maketitle

The modification of hadronic properties in a nuclear medium may be 
related  
to the partial restoration of chiral symmetry~\cite{Brown} -- an idea  
which is currently receiving considerable attention. 
Some experimental evidence for  
such effects has been discovered recently -- e.g.,
see Ref.~\cite{Cassing} for a review.

There is also a great deal of interest in possible 
signals of Quark-Gluon Plasma (QGP)
formation (or precursors to its formation) and $J/\Psi$ suppression 
is a promising candidate which has recently shown an anomalous 
result~\cite{Qm97,Vogt}.
On the other hand, there may be other mechanisms which produce an
increase in $J/\Psi$ absorption in a hot, dense medium. We are 
particularly interested in the rather exciting suggestion, based on the 
quark-meson coupling (QMC) model~\cite{Guichon}, that the charmed mesons,
$D$, $\bar{D}$, $D^*$ and $\bar{D}^*$, should suffer substantial 
changes in their 
properties in a nuclear medium~\cite{Tsushima}. This might be expected to
have a considerable impact on charm production in heavy ion collisions. 

In Ref.~\cite{Tsushima} it was found, for example, that at a density  
$\rho_B{=}3\rho_0 (\rho_0{=}0.15$~fm$^{-3}$) 
the $D$-meson would feel an attractive scalar 
potential of about $120$~MeV 
and an attractive vector potential
of about $250$~MeV. 
These potentials are comparable to
those felt by a $K^-$-meson~\cite{Kmeson,Kminus}, while the total potential 
felt by the $D$ is much larger than that for the vector mesons, 
$\rho$, $\omega$ and $\phi$~\cite{Saito}.  
Within QMC it is expected that the mass of 
the $J/\Psi$ should only be changed by a tiny amount in nuclear 
matter~\cite{Kmeson,Saito}. 
A similar result has also been obtained using QCD 
sum rules~\cite{Qcdsr}. 

In the light of these results, it seems that the charmed mesons,  
together with the $K^-$, are probably the best candidates to 
provide us with information on the partial restoration of 
chiral symmetry.  Both open charm production and the 
dissociation of charmonia in matter may therefore be used as new 
ways of detecting the modification of particle properties 
in a nuclear medium. 

The suppression of $J/\Psi$ production observed in relativistic 
heavy ion collisions, from $p{+}A$ up to central $S{+}U$ collisions,
has been well understood in terms of charmonium absorption in 
the nuclear medium. However, recent data from 
$Pb{+}Pb$ collisions show a considerably stronger $J/\Psi$ 
suppression~\cite{Qm97}. In an attempt to explain this 
``anomalous'' suppression of 
$J/\Psi$ production, many authors have studied one of two possible
mechanisms, namely 
hadronic processes~\cite{Brodsky,Capella,Martins,Mueller} 
and QGP formation~\cite{Matsui} 
(see Ref.~\cite{Vogt} for a review). 

In the hadronic dissociation scenario~\cite{Brodsky} it is well  
known that the $J/\Psi$ interacts with pions and $\rho$-mesons 
in matter, forming charmed mesons through the reactions,   
$\pi{+}J/\Psi {\to} D^*+\bar{D}, \bar{D}^*+D$ and 
$\rho{+}J/\Psi {\to} D+\bar{D}$.   
The absorption of $J/\Psi$ mesons on pions and $\rho$-mesons has been 
found to be important (see Refs.~\cite{Cassing,Capella,Gavin} and
references therein) in general and absolutely
necessary in order to fit the data on $J/\Psi$ production. 
Furthermore, the absorption on comovers should certainly play 
a more important role in $S+U$ and $Pb+Pb$ experiments, where hot, 
high density mesonic matter is expected to be achieved.
 
$J/\Psi$ dissociation on comovers, combined with the absorption on 
nucleons, is the main mechanism proposed as an alternative to 
that of Matsui and Satz~\cite{Matsui} -- namely the dissociation of the
$J/\Psi$ in a QGP. 
Note that both the hadronic and QGP scenarios predict $J/\Psi$
suppression but no mechanism has yet been found to separate them
experimentally.  

\begin{figure}[h]
\phantom{aa}\vspace{-12mm}\hspace{-2mm}
\psfig{file=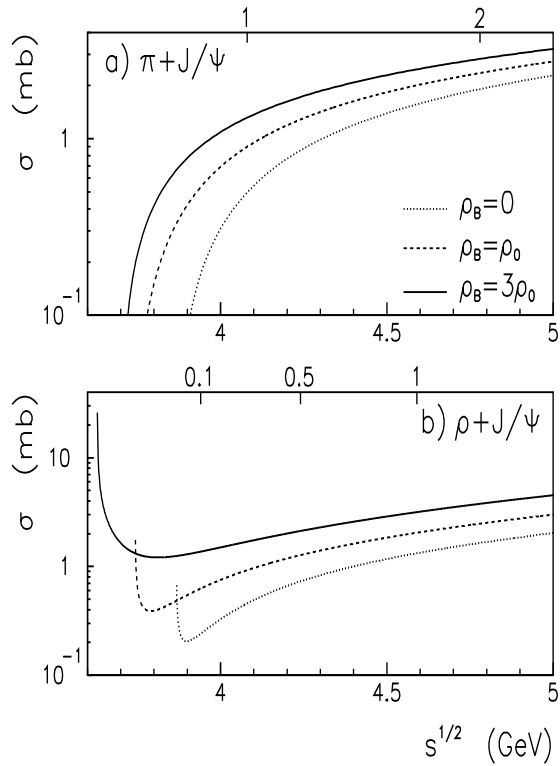,height=11.4cm,width=8.2cm}
\vspace{-17mm}
\caption[]{$\pi{+}J/\Psi$ (a) and $\rho{+}J/\Psi$ (b) dissociation
cross sections as functions of the invariant collision energy, $s^{1/2}$.
Results are shown for vacuum (the dotted line), $\rho_0$ (the dashed line) 
and $3\rho_0$ (the solid line).}
\label{char1}
\vspace{-2mm}
\end{figure}

Within the hadronic scenario the crucial point is the required 
dissociation strength. In particular, one needs a total cross section for 
the $\pi,\rho{+}J/\Psi$ interaction of around $1.5{\div}3$~mb
in order to explain
the data in heavy ion simulations~\cite{Cassing}. 
Recent calculations~\cite{Mueller} of the reactions 
$\pi{+}J/\Psi {\to} D+\bar{D}^*, \bar{D}+D^*$ and 
$\rho{+}J/\Psi {\to} D+\bar{D}$, based on $D$ exchange, 
indicate a  much lower cross section than this. 

The main uncertainty in the discussion of the 
$J/\Psi$ dissociation on a meson gas is given by the estimates 
of the $\pi,\rho{+}J/\Psi$ cross section~\cite{Mueller}. 
The predictions available for the  $\pi{+}J/\Psi$ cross section are given
in Refs.~\cite{Martins,Mueller}.
Following the meson exchange model 
of Ref.~\cite{Mueller}, we show by the dotted line in the  
Fig.~\ref{char1} the $\pi{+}J/\Psi$ (a) and $\rho{+}J/\Psi$ (b) 
dissociation cross sections calculated in free space.
Moreover, the upper axis of Fig.~\ref{char1}a) indicates the $\pi$-meson 
kinetic energy, $T_\pi$, given in the $J/\Psi$ rest frame,
which indicates that the pions should be sufficiently
hot to be above the $DD^\ast$ production threshold and to
dissociate the  $J/\Psi$-meson. By taking a thermal 
pion gas with average $T_\pi{\simeq}$150~MeV, one might conclude 
that independent of the $\pi{+}J/\Psi$ dissociation
model used~\cite{Martins,Mueller}, the rate of this
process is small. However, this situation changes when
the in-medium potentials of the charmed mesons are taken into account,
because they lower the $\pi{+}J/\Psi{\to}\bar{D}{+}D^\ast$ 
threshold. The upper axis of Fig.~\ref{char1}b)
shows the $\rho$-meson kinetic energy $T_\rho$ in the 
$J/\Psi$ rest frame. As was discussed in 
Refs.~\cite{Cassing,Vogt,Mueller}, the $J/\Psi$ dissociation 
might proceed on thermal $\rho$-mesons, because of the low
$\rho{+}J/\Psi{\to}\bar{D}{+}D^\ast$ reaction threshold.

As far as the meson properties in free space are concerned,
the Bethe-Salpeter (BS) and Dyson-Schwinger (DS) approaches
have been widely used~\cite{mir}. The application of BS approach
to the description of heavy-light quark systems
allows one to describe the D and B meson properties in free
space quite well~\cite{Weiss}. The DS approach at finite baryon density
was used~\cite{Blaschke} for the calculation of the in-medium
properties of $\rho$, $\omega$ and $\phi$ mesons.
The modification of the $\rho$ and $\omega$ meson masses
resulting from the DS equation is close to the calculations
with the  quark-meson coupling (QMC) model~\cite{Guichon,Tsushima,Saito}, 
while the $\phi$-meson mass reduction from Ref.~\cite{Blaschke}
is larger than the  QMC result.

Here, we use the quark-meson coupling model~\cite{Guichon},
which has been successfully applied not only to the problems
of nuclear binding and charge densities~\cite{Guichonf}
but also to the study of meson properties in a nuclear
medium~\cite{Tsushima,Kmeson,Saito}.
A detailed description of the Lagrangian density and the
mean-field equations
are given in Refs.~\cite{Tsushima,Kmeson,Saito,Guichonf}. 
The Dirac equations for the quarks and antiquarks in the $D$ and
$\bar{D}$ meson bags ($q$ stands for the light quarks hereafter), 
neglecting the Coulomb force,
are given by~\cite{Tsushima}:
\begin{eqnarray}
\left[ i \gamma \cdot \partial_x -
(m_q - V^q_\sigma)
\mp \gamma^0
\left( V^q_\omega +
\frac{1}{2} V^q_\rho
\right) \right] \nonumber \\
\times \left( \begin{array}{c} \psi_u(x)  \\
\psi_{\bar{u}}(x) \\ \end{array} \right)
 = 0,
\label{diracu}
\\
\left[ i \gamma \cdot \partial_x -
(m_q - V^q_\sigma)
\mp \gamma^0
\left( V^q_\omega
- \frac{1}{2} V^q_\rho
\right) \right] \nonumber \\  \times
\left(\begin{array}{c} \psi_d(x) \\ \psi_{\bar{d}}(x) \\ \end{array} \right)
 = 0,
\label{diracd}
\\
\left[ i \gamma \cdot \partial_x - m_{c} \right]
\psi_{c} (x)\,\, ({\rm or}\,\, \psi_{\bar{c}}(x)) = 0.
\label{diracsc}
\end{eqnarray}
The mean-field potentials for a bag in symmetric nuclear matter
are defined by $V^q_\sigma{=}g^q_\sigma
\sigma$,
$V^q_\omega{=}$ $g^q_\omega
\omega$ and
$V^q_\rho{=}g^q_\rho b$,
with $g^q_\sigma$, $g^q_\omega$ and
$g^q_\rho$ the corresponding quark and meson-field coupling
constants. 

The normalized, static solution for the ground state quarks or antiquarks
in the meson bags may be written as~\cite{Kmeson}:
\begin{eqnarray}
\psi_f (x) = N_f e^{- i \epsilon_f t / R_j^*}
\psi_f (\mbox{\boldmath $x$}),
\qquad (j = D, \bar{D}),
\label{wavefunction}
\end{eqnarray}
where $f{=}u$, $\bar{u}$, $d$, $\bar{d}$, $c$, $\bar{c}$
refers to quark flavors, and $N_f$ and $\psi_f(\mbox{\boldmath $x$})$
are the normalization factor and
corresponding spin and spatial part of the wave function. The bag
radius in medium, $R_j^*$, which generally depends on the hadron species to
which the quarks and antiquarks belong, is determined through the
stability condition for the  mass of the meson against the
variation of the bag radius~\cite{Tsushima,Saito,Guichonf}. 
The eigenenergies $\epsilon_f$ in
Eq.~(\ref{wavefunction}) in units of $1/R_j^*$  are given by
\begin{eqnarray}
\left( \begin{array}{c}
\epsilon_u \\
\epsilon_{\bar{u}}
\end{array} \right)
&=& \Omega_q^* \pm R_j^* \left(
V^q_\omega
+ \frac{1}{2} V^q_\rho \right),
\label{uenergy}
\\
\left( \begin{array}{c} \epsilon_d \\
\epsilon_{\bar{d}}
\end{array} \right)
&=& \Omega_q^* \pm R_j^* \left(
V^q_\omega
- \frac{1}{2} V^q_\rho \right),
\label{denergy}
\\
\epsilon_{c}
&=& \epsilon_{\bar{c}} =
\Omega_{c},
\label{cenergy}
\end{eqnarray}
where $\Omega_q^*
{=}\sqrt{x_q^2{+}(R_j^* m_q^*)^2}$, with
$m_q^*{=}m_q{-}g^q_\sigma \sigma$ and
$\Omega_{c}{=}\sqrt{x_{c}^2{+}(R_j^* m_{c})^2}$.
The bag eigenfrequencies, $x_q$ and $x_{c}$, are
determined by the usual, linear boundary condition~\cite{Guichonf}.

The $D$ and $\bar{D}$ meson masses
in symmetric nuclear matter are given by:
\begin{eqnarray}
m_D^* &=& \frac{\Omega_q^*
+ \Omega_c - z_D}{R_D^*}
+ {4\over 3}\pi R_D^{* 3} B,
\label{md}
\\
& &\left. \frac{\partial m_D^*}
{\partial R_D}\right|_{R_D = R_D^*} = 0.
\label{equil}
\end{eqnarray}
In Eq.~(\ref{md}), the $z_D$ parametrize the sum of the
center-of-mass and gluon fluctuation effects, and are assumed to be
independent of density~\cite{Guichonf}. The parameters are 
determined in free space to
reproduce their physical masses.

\begin{figure}[h]
\phantom{aa}\vspace{-10mm}\hspace{-1mm}
\psfig{file=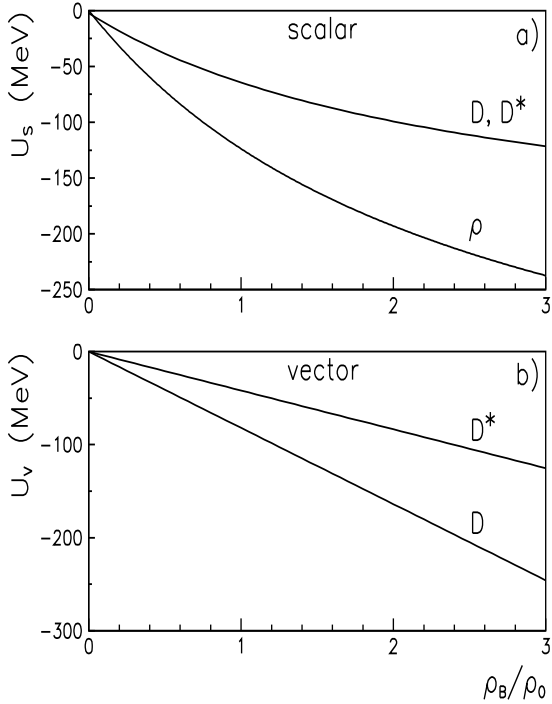,height=11.cm,width=8.1cm}
\vspace{-19mm}
\caption[]{The scalar (a) and vector (b) potentials for the 
$D$, $D^\ast$ and $\rho$ mesons, calculated for nuclear matter
as functions of the baryon density, in units of the
saturation density of nuclear matter, $\rho_0$=0.15 fm$^{-3}$.
Scalar potentials for $D$ and $D^\ast$ are indistinguishable.}
\label{char3}
\vspace{-4mm}
\end{figure}

In this study we chose the values $m_q{\equiv}m_u{=}m_d{=}5$
MeV and $m_c{=}1300$ MeV for the current quark masses, and $R_N{=}0.8$
fm for the bag radius of the nucleon in free space. Other input
parameters and some of the quantities calculated are given
in Refs.~\cite{Tsushima,Guichonf}.
We stress that while the model has a
number of parameters, only three of them, $g^q_\sigma$, $g^q_\omega$
and $g^q_\rho$, are adjusted to fit nuclear data -- namely the
saturation energy and density of symmetric nuclear matter and the bulk
symmetry energy. Exactly
the same coupling constants, $g^q_\sigma$, $g^q_\omega$ and
$g^q_\rho$, are used for the light quarks in the mesons as in the
nucleon.
Through Eqs.~(\ref{diracu}) -- (\ref{equil}) we self-consistently
calculate effective masses,
$m^*_D$,
and mean field potentials, $V^q_{\sigma,\omega,\rho}
$, in symmetric nuclear matter.
The scalar and vector potentials 
felt by the $D$ and $\bar{D}$ mesons
are given by~\cite{Tsushima}:
\begin{eqnarray}
U^{D^\pm}_s
&\equiv& U_s = m^*_D - m_D,
\label{spot}\\
U^{D^\pm}_v &=&
 \mp  (\tilde{V}^q_\omega - \frac{1}{2}V^q_\rho),
\label{vdpot1}
\end{eqnarray}
where, $\tilde{V}^q_\omega = 1.4^2 V^q_\omega$, which is assumed to 
be the same as that for the $K^+$ and $K^-$ 
mesons~\cite{Tsushima,Kmeson}. 
The isovector meson mean field potential, $V^q_\rho$,
is zero in symmetric nuclear matter.

Finally, the $D$, $D^\ast$ and $\rho$-meson potentials used in further 
calculations are shown in Fig.~\ref{char3} as a function of the
baryon density, in units of $\rho_0$=0.15 fm$^{-3}$.
Note that these potentials enter not only in the final state 
phase space (which becomes larger since the scalar masses
are reduced in matter), but also in the reaction amplitude and 
the initial $\rho$-meson mass. Furthermore, as observed earlier, 
the properties of the $J/\Psi$ meson are not 
significantly altered in medium within QMC.

Note that the total $D^-$-meson potential is repulsive, while the
$D^+$ potential is attractive, which is analogous to the
case for the $K^+$ and $K^-$ mesons, respectively~\cite{Tsushima,Kmeson}.
The threshold reduction is quite large when the nuclear density becomes 
large for the $D^+D^-$ pairs. Note that a similar situation holds for the
$K^+$ and $K^-$ production and, indeed, enhanced $K^-$-meson production
in heavy ion collisions, associated with the reduction of the
production threshold, has been partially confirmed
experimentally~\cite{Cassing,Kaos}.

\begin{figure}[h]
\phantom{aa}\vspace{-9mm}\hspace{-2mm}
\psfig{file=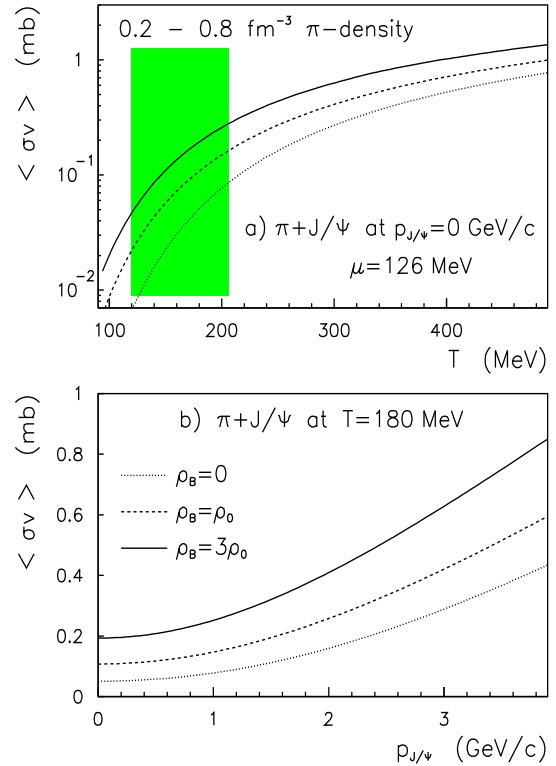,height=11.4cm,width=8.1cm}
\vspace{-15mm}
\caption[]{Thermally averaged $\pi{+}J/\Psi$ absorption
cross section as a function of the pion gas temperature, $T$,
and chemical potential, $\mu$, at $p_{J/\Psi}{=}0$
(a) and as a function of the
$J/\Psi$-momentum at $T{=}180$~MeV (b). The results are shown 
using the same notation as in Fig.1. The shadowed area indicates the
temperatures expected to be achieved in heavy ion collisions.}
\label{char2}
\vspace{-3mm}
\end{figure}

\begin{figure}[h]
\phantom{aa}\vspace{-8mm}\hspace{-2mm}
\psfig{file=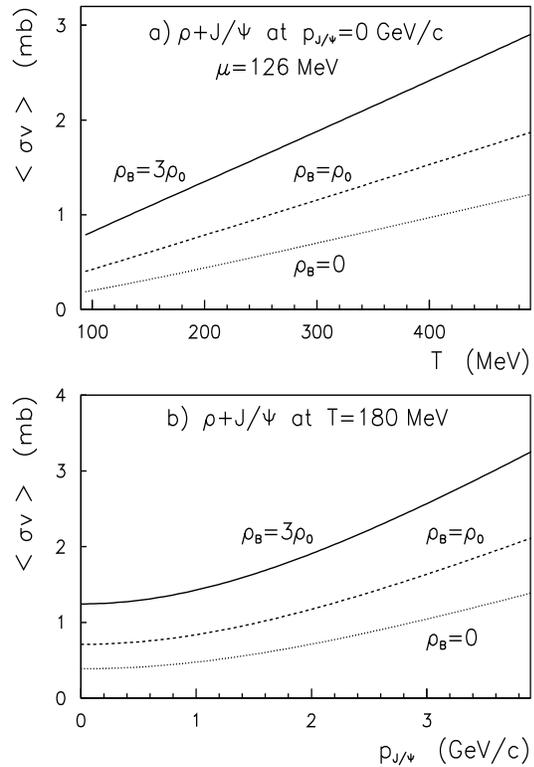,height=11.4cm,width=8.1cm}
\vspace{-16mm}
\caption[]{Thermally averaged $\rho{+}J/\Psi$ absorption
cross section as a function of the $\rho$-meson gas temperature $T$
with $p_{J/\Psi}{=}0$ (a) and as a function of the
$J/\Psi$-momentum at $T{=}180$~MeV (b). Notations are
similar to Fig.~\protect\ref{char2}.}
\label{char4}
\vspace{-3mm}
\end{figure}

We first discuss the thermally averaged
cross sections, ${\langle}{\sigma}v{\rangle}$,  for ${\pi{+}J/\Psi}$ 
and ${\rho{+}J/\Psi}$ dissociation in Figs.~\ref{char2} and \ref{char4}, 
when the free masses are used 
for the charmed mesons and the in-medium potentials are set to zero. 
They are shown by the dotted lines.
These results are needed for comparison with the calculations
including the potentials.

Since the pions are almost 
in thermal equilibrium, their energy spectrum is given by a Bose 
distribution with temperature, $T$, and chemical potential, 
$\mu$, where we have used the value, $\mu{=}126$~MeV~\cite{Kataja}.
The thermally averaged ${\pi{+}J/\Psi}$ 
cross section can be obtained by averaging over the
$\pi$-spectrum at fixed $J/\Psi$-momentum. The dotted line 
in Fig.~\ref{char2}a) shows ${\langle}{\sigma}{v}\rangle$ as a 
function of the pion gas temperature, $T$, which was calculated with 
zero $J/\Psi$ momentum relative to the pion gas.

The shadowed area in Fig.~\ref{char2}a) indicates the temperature range  
corresponding to the pion densities $0.2 - 0.8$~fm$^{-3}$, 
which are expected to be achieved in the heavy ion collisions presently
under consideration. 
In vacuum the $\pi{+}J/\Psi$ dissociation cross section 
is less than about $0.3$~mb. 
The thermally averaged absorption cross section for temperature, 
$T{=}180$~MeV, 
is shown in Fig.~\ref{char2}b) (the dotted line) as a
function of the $J/\Psi$ momentum.  
The thermally averaged cross section which we find, 
${\langle}{\sigma}v{\rangle}$, would be very difficult to detect 
with the present experimental capabilities.

A similar situation holds for the $\rho{+}J/\Psi$ dissociation,
as illustrated by the dotted line in Fig.~\ref{char4}. 
Indeed, the $J/\Psi$
absorption on comovers seems to be negligible~\cite{Mueller} 
in comparison with that needed to explain the $J/\Psi$ suppression, 
provided that we use the vacuum properties of the charmed mesons. 

On the other hand, this situation changes dramatically when we 
consider the effect of the vector and 
scalar potentials felt by the charmed $D$, $D^\ast$ and $\rho$ mesons  
as calculated by Eqs.~(\ref{spot}) and~(\ref{vdpot1}). 
The cross sections calculated for $\pi, \rho{+}J/\Psi$ collisions  
with the in-medium potentials are shown in Fig.~\ref{char1}, 
for densities, $\rho_0$ (the dashed line) and $3\rho_0$
(the solid line). The dotted line in Fig.~\ref{char1}
indicates  the free space cross sections.

Clearly the $J/\Psi$ absorption cross sections are substantially enhanced
for both the $\pi{+}J/\Psi$ and $\rho{+}J/\Psi$ reactions, not only 
because of the downward shift of the reaction threshold, but also 
because of the in-medium effect on the reaction amplitude. 
Moreover, now the $J/\Psi$ absorption
on comovers becomes  density dependent -- a 
crucial finding given the situation in actual heavy ion collisions.
These effects have never been considered before. 
The comover absorption cross section is calculated as
a function of baryon density for the first time.

The thermally averaged, in-medium $\pi{+}J/\Psi$ and $\rho{+}J/\Psi$
absorption cross sections, ${\langle}{\sigma}v{\rangle}$, 
are shown by the dashed and solid lines in Figs.~\ref{char2} and 
\ref{char4}, respectively. We find that  
${\langle}{\sigma}v{\rangle}$ depends 
very strongly on the nuclear density. Even for
$p_{J/\Psi}{=}0$, with a pion gas 
temperature of 120~MeV, which is close to the saturation pion density, 
the thermally averaged $J/\Psi$ absorption cross section on the 
pion, at $\rho_B{=}3\rho_0$, is about a factor of 7 larger than that 
at $\rho_B{=}0$ (i.e., with no effect of the in-medium modification 
-- see Fig.~\ref{char2}a)).  

The thermally averaged $\rho{+}J/\Psi$ dissociation cross section  
at $\rho_B{=}3\rho_0$ 
becomes larger than 1~mb. Thus, the $J/\Psi$ absorption on $\rho$-mesons
should be appreciable, even though the $\rho$-meson density  
is expected to be small in heavy ion collisions.  
We note that dynamical calculations~\cite{Cassing}
suggest that the $\rho$-density should be around half of the pion density
in $Pb{+}Pb$ collisions. 

\begin{figure}[h]
\phantom{aa}\vspace{-10mm}\hspace{-2mm}
\psfig{file=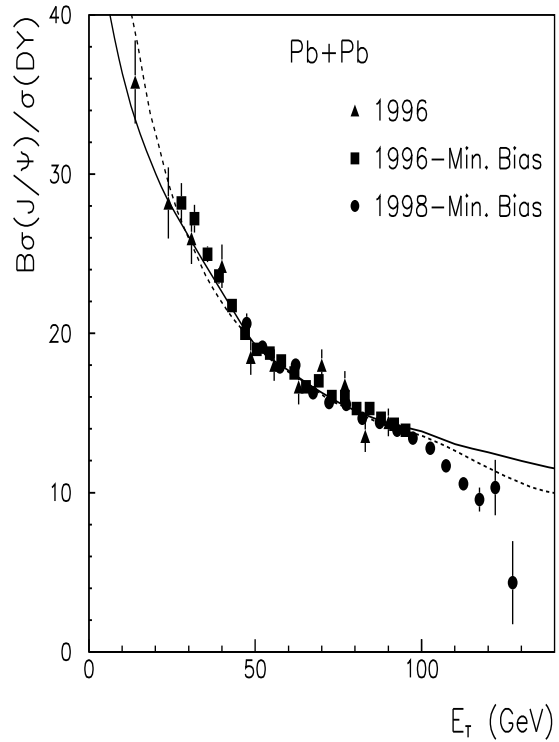,height=11.3cm,width=8.2cm}
\vspace{-17mm}
\caption[]{The ratio of the $J/\Psi$ over Drell-Yan cross sections from 
$Pb{+}Pb$ collisions as function of the transverse energy $E_T$. 
Data are from Ref.~\cite{Qm97,Last}. The solid line shows our calculations
with the density dependent cross section for
$J/\Psi$ absorption on comovers. The dashed line indicates the 
calculations with phenomenological cross section  
${\langle}{\sigma}v{\rangle}$$\simeq$1~mb~\protect\cite{Capella1}.
For both calculations the nuclear absorption cross section was
taken as 4.5~mb.}
\label{char7}
\vspace{-3mm}
\end{figure}

In order to compare our results with the NA38/NA50 
data~\cite{Qm97,Last} on $J/\Psi$ suppression in $Pb{+}Pb$
collisions, we have adopted the heavy ion model proposed in
Ref.~\cite{Capella} with the $E_T$ model from 
Ref.~\cite{Capella1}. We  introduce the absorption
cross section on comovers as function of the density of comovers,
while the nuclear absorption cross section is taken 
as 4.5~mb~\cite{Capella1}. 
Our calculations are shown in Fig.~\ref{char7}.
The dashed line in Fig.~\ref{char7} shows the calculations 
with the phenomenological constant cross section for $J/\Psi$ 
absorption on comovers ${\langle}{\sigma}v{\rangle}$$\simeq$1~mb 
and is identical to the results given in Ref.~\cite{Capella1}.
The solid line in 
Fig.~\ref{char7} shows the calculations with the
density dependent cross section ${\langle}{\sigma}v{\rangle}$
for $J/\Psi$ absorption 
on comovers calculated in this work. Both curves clearly reproduce  
the data~\cite{Qm97} quite well, 
including most recent results from NA50 on the ratio
of $J/\Psi$ over Drell-Yan cross sections, as a function
of the transverse energy up to $E_T$=100~GeV. 
It is important to note that if one neglected the in-medium modification of 
the $J/\Psi$ absorption cross section the large cross section
${\langle}{\sigma}v{\rangle}$$\simeq$1~mb cannot be justified by 
microscopic theoretical calculations and thus
the NA50 data~\cite{Qm97,Last} cannot be described. 
 
Furthermore, we notice
that our calculations with in-medium modified absorption
provide a significant improvement in the understanding of the 
data~\cite{Qm97} compared to the models quoted by NA50~\cite{Last}.
The basic difference between our results and those quoted by
NA50~\cite{Last} is that in previous 
heavy ion calculations~\cite{Cassing,Vogt,Capella} the cross section for 
$J/\Psi$ absorption on comovers was taken as a free parameter to
be adjusted to the data~\cite{Qm97,Last} and was never
motivated theoretically.

To summarize, we have studied $J/\Psi$ dissociation in a gas of $\pi$ and
$\rho$ mesons, taking into account for the first time the density
dependence of the scalar and vector potentials which 
the mesons feel in nuclear matter. 
We found a substantial density dependence of the  $J/\Psi$ absorption 
rate as a result of the changes in the properties of the charmed mesons
in-medium. This aspect has 
never been considered before when analyzing $J/\Psi$ production in 
heavy ion collisions.

Moreover, when we introduce the density dependent cross section 
on comovers into a heavy ion calculation and compare our results 
with most recent NA50 data~\cite{Qm97,Last}, our calculations indicate
very good agreement with the NA50 data up to transverse energy of order 
100~GeV.  

A.S and K.S would like to acknowledge the warm hospitality at the CSSM, 
where this work was initiated and completed. 
A.S. was supported in part by the Forschungzentrum J\"{u}lich.
K.S. would also like to acknowledge the support of
the Japan Society for the Promotion of Science.
This work was supported by the Australian Research Council and the 
University of Adelaide.
A.S. would also like to acknowledge discussions with D. Kharzeev.


\begin{thebibliography}{99}
\bibitem{Brown}
        G.E. Brown and M. Rho, Phys. Rev. Lett. 66 (1991) 2720; 
        T. Hatsuda and S.H. Lee, Phys. Rev. C 46 (1992) 34.        
\bibitem{Cassing}
        W. Cassing and E. Bratkovskaya, Phys. Rep. 308 (1999) 65.
\bibitem{Qm97} Quark Matter '97, Nucl. Phys. A 638 (1998);
        M. C. Abreu et al. (NA50 Collaboration),
        Phys. Lett. B 410  (1997) 337; 
        Phys. Lett. B   450 (1999) 456.
\bibitem{Vogt}
        R. Vogt, Phys. Rep. 310 (1999) 197. 
\bibitem{Guichon} 
        P.A.M. Guichon, Phys. Lett. B 200 (1989) 235. 
\bibitem{Tsushima}
        K. Tsushima, D.H. Lu, A.W. Thomas, K. Saito and R.H. Landau,
        Phys. Rev.  C 59  (1999) 2824;
        A. Sibirtsev, K. Tsushima and A.W. Thomas, Eur. Phys. J. A
        6 (1999) 351.
\bibitem{Kmeson}K. Tsushima, K. Saito, A.W. Thomas and S.W. Wright,
        Phys. Lett.  B 429  (1998) 239;
        {\it ibid.}  B 436 (1998), 453.
\bibitem{Kminus}T. Waas and W. Weise, Nucl. Phys. A 625 (1997) 287;
        T. Waas, N. Kaiser and W. Weise, Phys. Lett. B 379 (1996) 34;
        A. Sibirtsev and W. Cassing, Nucl. Phys. A 641 (1998) 476.  
\bibitem{Saito}
        K. Saito and A.W. Thomas, Phys. Rev. C  51 (1995) 2757;
        K. Saito, K. Tsushima and A.W. Thomas, Phys. Rev. C  55 (1997) 2637;
        K. Tsushima, D.H. Lu, A.W. Thomas and K. Saito,
        Phys. Lett. B 443 (1998) 26. 
\bibitem{Qcdsr} 
        F. Klingl et al., Phys. Rev. Lett.  82 (1999) 3396; 
        A. Hayashigaki, Prog. Theor. Phys. 101 (1999) 923. 
\bibitem{Brodsky}
        S.J. Brodsky and A.H. Mueller, Phys. Lett. B 206 (1988) 685;
        S. Gavin, M. Gyulassy and A. Jackson, 
        Phys. Lett. B 207  (1988) 257; 
        R. Vogt, M. Prakash, P. Koch and T.H. Hansson, 
        Phys. Lett. B  207  (1988) 264;
        J.-P. Blaizot and J.-Y. Ollitrault, 
        Phys. Rev. D  39  (1989) 232;
        C.-Y. Wong, E.S. Swanson and T. Barnes, hep-ph/9912431;
        nucl-th/0002034.
\bibitem{Capella}
        N. Armesto and  A. Capella, J. Phys. G {  23}, 1969 (1997); 
        Phys. Lett. B  430  (1998) 23; 
        N. Armesto, A. Capella and E.G. Ferreiro, 
        Phys. Rev. C  59  (1999) 395. 
\bibitem{Capella1}
        A. Capella, E.G. Ferreiro and A.B. Kaidalov,
        hep-ph/0002300.
\bibitem{Martins}
        D. Kharzeev and H. Satz, Phys. Lett. B 334 (1994) 155;
        K. Martins, D. Blaschke and E. Quack,
        Phys. Rev. C 51 (1995)  2723;
        D. Kharzeev, H. Satz, A. Syamtomov and G. Zinovev,
        Phys. Lett. B 389 (1996) 595.
\bibitem{Mueller}
        S.G. Matinyan and B. M\"uller,
        Phys. Rev. C  58  (1998) 2994;
        B. M\"uller, Nucl. Phys. A 661 (1999) 272.
\bibitem{Matsui}
        T. Matsui and H. Satz, Phys. Lett. B 178 (1986) 416. 
\bibitem{Gavin}
        S. Gavin and R. Vogt, 
        Nucl. Phys. B 345 (1990) 104;
        S. Gavin, H. Satz, R.L. Thews and R. Vogt,
        Z. Phys. C 61 (1994) 351.
\bibitem{mir}
        V.A. Miransky, 
        {\it Dynamical Symmetry Breaking in Quantum Field Theories},
        World Scientific Publishing Co., (1993).
\bibitem{Weiss}
        Yu.L. Kalinovsky and C. Weiss, Z. Phys.  C 63 
        (1994) 275.
\bibitem{Blaschke}
        D. Blaschke and C.D. Roberts,  Nucl. Phys.  A 642
        (1998) 197.
\bibitem{Guichonf}
        P.A.M. Guichon, K. Saito, E. Rodionov and A.W. Thomas, 
        Nucl. Phys.  A 601 (1996) 349;
        K. Saito, K. Tsushima and A.W. Thomas, 
        Nucl. Phys. A 609 (1996) 339.
\bibitem{Kaos}
        F. Laue et al., Phys. Rev. Lett. 82
        (1999) 1640;
        A. Schroter et al., Z. Phys. A 350  (1994) 101;
        J.L. Ritman et al., Z. Phys. A 352 (1995) 355;
        R. Barth et al., Phys. Rev. Lett.  78 (1997) 4007;
        Y. Shin  et al., Phys. Rev. Lett.  81 (1998) 1576;
        G.Q. Li, C.M.  Ko and  X.S. Fang, Phys. Lett.  B 329
        (1994) 149;
        W. Cassing, E.L. Bratkovskaya, U. Mosel, S. Teis and
        A. Sibirtsev,  Nucl. Phys. A 614 (1997) 415;
        G.Q Li, C.M. Ko and G.E. Brown, Phys. Lett.
        B 381 (1996) 17.
\bibitem{Kataja}
        K. Kataja and P.V. Ruuskanen, Phys. Lett. B 243 (1990) 181.
\bibitem{Last}
        M. C. Abreu et al. (NA50 Collaboration),
        Phys. Lett. B  477 (2000) 28.
\end{thebibliography}
\end{document}